%\magnification=\magstep1

\def\mj{$\,M_J\,$}

\def\Dwa{$\,$\uppercase\expandafter{\romannumeral5}$\,$}
 
\def\sles{\lower2pt\hbox{$\buildrel {\scriptstyle <}
   \over {\scriptstyle\sim}$}}
\def\sgreat{\lower2pt\hbox{$\buildrel {\scriptstyle >}
   \over {\scriptstyle\sim}$}}
\def\lapprox{\lower2pt\hbox{$\buildrel \lower2pt\hbox{${\scriptstyle<}$}
   \over {\scriptstyle\approx}$}}
\def\gapprox{\lower2pt\hbox{$\buildrel \lower2pt\hbox{${\scriptstyle>}$}
   \over {\scriptstyle\approx}$}}

\def\gta{\lower 0.8ex\hbox{$\buildrel > \over \sim\ $}}
\def\lta{\lower 0.8ex\hbox{$\buildrel < \over \sim\ $}}

\newcount\sss
\sss=0
\def\super{\advance\sss by 1 $\!^{\number\sss}$}

\font\tinyrm=cmr8 at 8truept
\font\tinybold=cmbx8 at 8truept
\font\tinyit=cmti8 at 8truept
\def\tiny{\let\rm=\tinyrm\let\bf=\tinybold\let\it=\tinyit}

\font\smallrm=cmr10 at 10truept
\font\smallbold=cmbx10 at 10truept
\font\smallit=cmti10 at 10truept
\def\small{\let\rm=\smallrm\let\bf=\smallbold\let\it=\smallit}

\font\normalrm=cmr12 at 12truept
\font\normalbold=cmbx12 at 12truept
\font\normalit=cmti12 at 12truept
\def\normal{\let\rm=\normalrm\let\bf=\normalbold\let\it=\normalit}

\font\bigrm=cmr10 scaled 1440
\font\bigbold=cmbx10 scaled 1440
\font\bigit=cmti10 scaled 1440
\def\big{\let\rm=\bigrm\let\bf=\bigbold\let\it=\bigit}
\baselineskip 24truept

\centerline{{\bigbold 
Atmospheric, Evolutionary, and Spectral Models}}
\centerline{{\bigbold of the 
Brown Dwarf Gliese 229 B
}}
\bigskip
\leftline{{\bf M.S. Marley$^{\ast}$, D. Saumon$^{\dag}$, T. Guillot$^{\dag}$, R.S.
Freedman$^{\ddag}$, W.B. Hubbard$^{\dag}$, A. Burrows$^{\S}$,}}

\leftline{{\bf \& J.I. Lunine$^{\dag}$}}

\vskip 1cm
\leftline{$^{\ast}$Department of Astronomy, New Mexico State University,
Box 30001/Dept. 4500, Las Cruces NM 88003}
\leftline{$^{\dag}$Lunar \& Planetary Laboratory, University of Arizona, Tucson AZ 85721}
\leftline{$^{\ddag}$Sterling Software, NASA Ames Research Center, Moffett Field CA 94035}
\leftline{$^{\S}$Departments of Physics and Astronomy, University of Arizona, Tucson AZ 
85721}

\baselineskip=24truept
\vskip 1cm

{\bf  Theoretical spectra and evolutionary models that span the giant planet--brown dwarf
continuum have been computed based on the recent discovery of the brown dwarf, Gliese 229 B.  
A flux enhancement in the 4--5 micron window is a universal feature from Jovian planets
to brown dwarfs.  We confirm the existence of methane and water in Gl 229 B's spectrum and find
its mass to be 30 to 55 Jovian masses.  Although these calculations focus on Gliese 229 B, they are 
also meant to guide future searches for extra-solar giant planets and brown dwarfs.}

\vfill\eject

Brown dwarfs inhabit a realm intermediate between the more massive stars and
the less massive planets.  Their thermal infrared emission 
is powered by the release of gravitational potential energy as regulated
by their atmospheres.  Long known only as theoretical constructs,
the discovery of the first unimpeachable brown dwarf (1,2) allows
a detailed study of a representative of this population of objects.
Gliese 229 B, the recently-discovered companion to Gliese (Gl) 229 A,
has an estimated luminosity of $6.4 \pm 0.6 \times 10^{-6} 
L_{\odot}$ (solar luminosity), an effective temperature, $T_{\rm eff}$,
below 1200 K, and a clear signature of methane in its spectrum (3).
Since there can be no stars cooler than 1700 K, with luminosities below 
$5 \times 10^{-5} L_{\odot}$, or
with methane bands (4), Gl 229 B's status as one of the long-sought brown dwarfs is now beyond question.
However, models of Gl 229 B's atmosphere and evolution are required to derive 
its physical properties
and the previous lack of observations had inhibited the generation of theoretical spectra.
To remedy this, we coupled model spectra and evolutionary calculations 
to estimate the object's $T_{\rm eff}$, $L$, surface gravity $g$, mass
$M$, radius $R$, and age $t$, and to find useful spectral diagnostics.  
The recent discoveries of planets 51 Pegasi B, 
70 Virginis B, 47 Ursae Majoris B,  and Gl 411 B (5) have doubled
the number of known Jovian planets.  There is now an extraordinarily rich
variety of low-temperature, low-mass 
(0.3 -- 84 $M_{\rm J}$ (Jupiter mass)) planets and brown dwarfs. 
Our improved evolutionary models and spectra, here applied to Gl 229 B, are meant
to facilitate the study and interpretation of these objects. 

To compute the atmospheric temperature profile for brown dwarfs
in the relevant temperature range (600--1200 K), we adapt a
model originally constructed to study the atmospheres
of the Jovian planets and Titan (6).  
We assume a standard solar composition for the bulk of
the atmosphere (7).
Refractory elements (for example Fe, Ti, and silicates) condense deep in the
atmosphere for $T_{\rm eff} \approx 1000$ K, and thus
have negligible gas-phase abundance near the photosphere, as
is also true in the atmosphere of Jupiter ($T_{\rm eff} = 124$ K).
For an atmosphere similar to that of Gl 229 B, chemical equilibrium
calculations indicate that C, N, O, S, and P
are found mainly in the form of methane (CH$_4$), ammonia (NH$_3$), 
water (H$_2$O), hydrogen sulfide (H$_2$S), and phosphine (PH$_3$),
respectively.   However, deep in the atmosphere, chemical equilibrium
favors CO over CH$_4$ and $\rm N_2$ over $\rm NH_3$. 
Our model atmosphere incorporates 
opacities of these molecules, H$_2$, and He (8) in their respective
solar abundances and includes no other elements.  

To constrain the properties of Gl 229 B, we construct a grid of brown dwarf model
atmospheres with $T_{\rm eff}$ ranging from 600 to 1200 K and 
$100<g<3200$ m s$^{-2}$.  For each case we compute a self-consistent
radiative-convective equilibrium temperature profile and the emergent radiative
flux (9).
Absorption of radiation from Gl 229 A is included in our model, but contributes
negligibly to Gl 229 B's energy balance owing to the large orbital
separation ($\ge 44$ AU) and faintness of Gl 229 A.

Emergent spectra of brown dwarf atmosphere models compared to 
observed fluxes (Fig.~1) (1,10) show the influence of a minimum in the molecular
opacities at wavelengths around $4-5$ $\mu$m.   As in the case for
Jupiter, this minimum allows radiation to escape from deep,
warm regions of the atmosphere.  Clearly, this wavelength region
is advantageous for future brown dwarf searches.  By comparison,
the widely-used K band at 2.2 $\mu$m is greatly suppressed by strong CH$_4$ and
H$_2-$H$_2$ absorption features.
Beyond 13 $\mu$m, the decreasing flux falls slightly more rapidly than  a 
Planck distribution with a brightness temperature near 600 K.

Our computed spectra (Fig. 1) are a good match with the data in 
the $1.2-1.8$ $\mu$m window regions, but deviate
at 1 $\mu$m, in the window centered on 2.1 $\mu$m, and in regions of low flux.
Our best-fitting models reproduce the observed broad band fluxes (3) reasonably well.
While many individual spectral features of $\rm CH_4$ and $\rm H_2O$ are
reproduced, particularly near $1.7$ and $2.0\,\rm\mu m$, the overall band shapes are not well accounted 
for in the 1 -- 2.5 $\mu$m
region (Fig.~1a).  We attribute these discrepancies to a poor knowledge of
the CH$_4$ opacity and, to a lesser extent, the H$_2$O opacity.  
Although we have combined several sources of varying accuracy (8)
to generate as complete a description of the CH$_4$ opacity as possible, methane  
line lists are based on laboratory measurements at room temperature and
do not include lines from higher energy levels that would be populated
at brown dwarf temperatures.
Thus, the opacity of $\rm CH_4$ at $T \approx 1000\,$K is the most likely
cause of the
mismatches seen in the 1.6 -- 1.8 $\mu$m band and at $\lambda > 2.1\,\rm \mu m$.

Clouds may alter the atmospheric
structure and spectrum of Gl 229 B, as they do in the atmospheres of
planets of our solar system. 
Extrapolating from results for Jupiter (11) and using
more recent chemical equilibrium calculations (12),
we find that the following additional molecules are expected to condense 
between $10^{-3}$ and
10 bars: NH$_4$H$_2$PO$_4$, ZnS, K$_2$S, Na$_2$S, and MnS.
If a relatively large proportion of condensed particles is retained in
the atmosphere, cloud layers could affect the
structure of the brown dwarf, making it hotter by as much as 100 K at 1
bar (depending on the uncertain particle sizes and optical properties).
Clouds might increase the flux in the K band, due to the higher temperatures,
and lower the flux below 1.3 $\mu$m, due to scattering.

    Given these uncertainties, our best fits for the bolometric luminosity, the
observed spectrum, and the photometry give combinations of
$T_{\rm eff}$ and $g$ lying in the range $850 < T_{\rm eff}\,{\rm (K)} < 1100$
and $g < 3000\,$m$\,$s$^{-2}$ (Fig. 2).  Lower $T_{\rm eff}$ are allowed for 
$g < 300\,$m$\,$s$^{-2}$, but the shapes of the J and H bands increasingly 
deviate from the 
observations.  The high-$T_{\rm eff}$ limit arises from the inability to fit 
simultaneously the bolometric luminosity and the 10$\,\mu$m flux.

A determination of  Gl 229 B's gravity via spectral matching would impose a direct  
constraint on its mass.  Although $g$ is a function of both mass
and radius, the radii of brown dwarfs in this temperature range
vary relatively little as the mass varies by an order
of magnitude.  However, at the present stage of the analysis 
the gravity is poorly constrained since high $g$, high $T_{\rm eff}$
models fit the spectra as well as lower $g$ and $T_{\rm eff}$ ones.
The model spectra suggest that high spectral resolution ($\lambda/\Delta \lambda \gta 1000$)
observations at $1.8$--$2.1\,\rm\mu m$ may provide a tighter constraint
on $g$.  

The depth at which the atmosphere becomes convective depends
upon the specified model gravity and effective temperature.
At the highest-pressure point of each model
atmosphere, where the temperature-pressure profile merges with
an adiabat, the interior entropy is calculated for the purpose of matching
an interior temperature distribution to the given values of $(T_{\rm eff}, g)$.  The
full evolutionary behavior of a brown dwarf
is obtained by supplementing previous
boundary conditions for objects with masses $\sim 0.3 - 15 \, M_J$ (Jupiter
mass units) (13, 14) with our grid of nongrey
model atmospheres.  Such evolutionary models are needed because 
$R$ varies with mass and age by up to 30\%.  The precise radius of
the object is important because we must match not only Gl 229 B's spectrum, but
also the inferred bolometric luminosity: $L=4 \pi R^2 \sigma T_{\rm eff}^4$
($\sigma$ is the Stefan-Boltzmann constant).  Our results can be summarized by the following
approximate fitting formulas ($g$ in m s$^{-2}$, $T_{\rm eff}$ in K):
$$
M= 36 \, M_J \, \, (g/1000)^{0.64} \, (T_{\rm eff}/1000)^{0.23},  \eqno(1)
$$
$$
t= 1.1 \, {\rm Gyr} \, \, (g/1000)^{1.7} \, (T_{\rm eff}/1000)^{-2.8},  \eqno(2)
$$
$$
R= 67200 \, {\rm km} \, \, (g/1000)^{-0.18} \, (T_{\rm eff}/1000)^{0.11}.  \eqno(3)
$$

The effective temperature and surface gravity of Gl 229 B can now be constrained  by 
three sets of 
observations (which are not independent of each other): (i)
the observed spectrum from 1 to 2.5 $\mu$m; (ii) the broadband flux in several bandpasses 
from 2 to 13 $\mu$m (10); and (iii) the bolometric luminosity of the object (3).
These constraints then limit $g<2200\,\rm m\,s^{-2}$ and $T_{\rm eff} = 960 \pm 70\,\rm K$ 
(Fig.~2). Since the reported age of Gl 229 A is $\gta 1\,\rm Gyr$ (1),
$g$ is further constrained to lie in the range 800 to $2200\,\rm m\,sec^{-2}$ (Fig.~3).

In the 
atmospheres of Gl 229 B and Jupiter, convection commences as the optical depth to
thermal photons becomes large, and the temperature profile closely approaches an
adiabatic profile at deeper levels owing to efficient convection (Fig.~4).  
In some models, particularly the lower gravity models and those
with $T_{\rm eff}< 900$ K,
the radiative-equilibrium lapse rate exceeds the adiabatic
lapse rate over a several-bar region near 1 bar.  These
atmospheres exhibit two convective regions, a lower region,
presumably continuing to great depth,
and an upper, detached convective zone.  Such a detached convective
zone is also predicted for the atmosphere of Jupiter (15).

A stellar evolution code and atmosphere models have allowed us to
estimate the physical properties of the brown dwarf, Gl 229 B.   
We derive an effective temperature of
$960 \pm 70$ K and a gravity between 800 and 2200 m s$^{-2}$.
These results translate into masses and ages of 30--55 \mj and 1--5 Gyr, respectively.
As Eq 1 and Fig. 3 indicate, gravity maps almost directly
into mass, and ambiguity in the former results in uncertainty in the latter.
Since the inferred mass of Gl 229 B exceeds that required for deuterium burning (14),
deuterium-bearing molecules should not be present in its atmosphere.
While the near infrared spectrum of Gl 229 B is dominated by $\rm H_2O$,
we confirm the presence of $\rm CH_4$ in the atmosphere from our modeling of its features at
1.6--1.8 $\mu$m, 2.2--2.4 $\mu$m, and 3.2--3.6 $\mu$m.  
In addition, we find a flux enhancement in 
the window at 4--5 $\mu$m throughout the T$_{\rm eff}$ range from 124 K (Jupiter)
through 1300 K, and, hence, that this band is a universal diagnostic for brown dwarfs 
and planets.  

\vfill\eject

\centerline{\bf References and Notes}
\bigskip

\item{1.}Nakajima, T. {\it et al.} {\it Nature} {\bf 378}, 463 (1995).

\item{2.}Oppenheimer, B.R., Kulkarni, S.R., Matthews, K., \& Nakajima, T. {\it Science} {\bf 270},
1478 (1995).

\item{3.} Matthews, K., Nakajima, T., Kulkarni, S.R. and Oppenheimer B.R. {\it Astrophys. J.},
submitted. Geballe, T.R., Kulkarni, S.R., Woodward, C.E., and Sloan, G.C. {\it Astrophys. J. Lett.},
in press.

\item{4.}Burrows, A., Hubbard, W. B., Saumon, D., \& Lunine, J. I.\ {\it Astrophys.\ J.} {\bf 406}, 158 (1993);
Lunine, J.I., Hubbard, W.B. and Marley, M.S. {\it Astrophys.\ J.}, {\bf 310}, 238 (1986).

\item{5.}Mayor, M. \& Queloz, D. {\it Nature}, {\bf 378}, 355 (1995);
Marcy, G.W., \& Butler, R.P. {\it Astrophys.\ J.\ Lett.}, submitted (1996);
Butler, R.P. \& Marcy, G.W. {\it Astrophys.\ J.\ Lett.}, submitted (1996);
Gatewood, G. {\it BAAS}, in press.

\item{6.}McKay, C.P, Pollack, J.B., \& Courtin, R. {\it Icarus} {\bf 80}, 23 (1989);
Marley, M.S., McKay, C.P., \&  Pollack, J.B. {\it Icarus}, submitted.

\item{7.}Anders, E., \& Grevesse, N. {\it Geochim. Cosmochim. Acta} {\bf 53},
197 (1989).

\item{8.} The opacity calculations include collision-induced absorption by H$_2$-H$_2$
[Borysow, A. \& Frommhold, L. {\it Astrophys. J.} {\bf 348}, L41 (1990)]
and H$_2$-He [Zheng, C. \& Borysow, A. {\it Icarus} {\bf 113}, 84 (1995)] and
references therein,
free-free absorption by H$_2^{-}$
[Bell, K.L. {\it J.  Phys. B} {\bf 13}, 1859 (1980)],
bound-free absorption by H$^{-}$
[John, T.L. {\it Astron. \& Astrophys.} {\bf 193}, 189 (1988)],
and Rayleigh scattering. 
The absorptions of NH$_3$, CH$_4$, and PH$_3$ were calculated using the HITRAN
data base [Hilico, J.C., Loete, M., \& Brown, L.R., Jr. {\it J. of Mol.
Spectr.} {\bf 152}, 229 (1992)] with corrections and extensions.  
Additional tabulations [Strong, K., Taylor, F.W., Calcutt, S.B., Remedios, J.J., 
\& Ballard, J. {\it J. Quant. Spectr. Radiat. Transfer \bf 50}, 363 (1993)]
were used where necessary for $\rm CH_4$, especially shortwards of $1.6\,\rm \mu m$.  Data for
H$_2$O and H$_2$S were computed from a direct numerical diagonalization 
[Wattson, R.B., and Rothman L.S. {\it JQSRT} {\bf 48}, 763 (1992)]
by R.B. Wattson (personal communication).  Absorption by CO [Pollack {\it et al. 
Icarus \bf 103}, 1 (1993)] and PH$_3$ opacity was included in the
spectral models, but not in the temperature profile computation.
The baseline models assume the atmosphere to be free of clouds.

\item {9.}
For the temperature profile computation, molecular opacity was treated using 
the k-coefficient method
[Goody, R., West, R., Chen, L., \& Crisp, D.
{\sl J. Quant. Spectr. Radiat. Transfer \bf 42}, 539 (1989)].
After a radiative-equilibrium temperature profile was found, the
atmosphere was iteratively adjusted to self-consistently solve
for the size of the convection zones, given the specified internal
heat flux.  Given the radiative-convective temperature-pressure profiles,
high-resolution synthetic spectra were generated by solving the
radiative transfer equation
[Bergeron, P., Wesemael, F., and Fontaine, G. {\it Ap.J.} {\bf 367}, 253 (1991)]
Eighteen thousand frequency points were used in the 1 -- 15.4$\, \mu$m spectral
region.  These spectra were smoothed with a Gaussian-bandpass
filter giving a final resolution of $\lambda / \Delta \lambda =600$.

\item{10.}Matthews, K., Nakajima, T., Kulkarni, S., \& Oppenheimer, B. {\it IAU Circ. \#6280} (1995).

\item{11.}Fegley, B., Jr \& Lodders, K. {\it Icarus} {\bf 110}, 117 (1994).

\item{12.}K. Lodders (personal communication).

\item {13.} Burrows, A., Saumon, D., Guillot, T., Hubbard, W.B., \& Lunine, J.I.
{\it Nature} {\bf 375}, 299 (1995).

\item{14.}Saumon, D., {\it et al.} {\it Astrophys.\ J. \bf 460},  993 (1996).

\item{15.} Guillot, T., Gautier, D., Chabrier, G., \& Mosser, B. {\it Icarus} {\bf 112}, 337 (1994).

\item{16.} Lindal, G. {\it Astron.J.} {\bf 103}, 967 (1992).

\item {17.} D.S. is a Hubble Fellow.  T.G. is supported by the European Space
Agency.  This research was supported by grants from the National
Aeronautics \& Space Administration, and from the National Science Foundation.
We thank T. Geballe for digital versions of the Gl 229 B spectrum, K. Lodders
for chemical-equilibrium calculations, and K. Zahnle for an insightful
review.

\vskip 1cm
\vfill\eject
{\bf Figure Captions}
\medskip

\item{\bf Fig. 1}  A): Synthetic spectra for (bottom to
top) $T_{\rm eff}$ = 890, 960, 1030 K and $g = 1000$ m s$^{-2}$, together
with data (3) (red line).
The three curves in B) are calculated for the same values of
$T_{\rm eff}$ and $g$; colored
boxes show the photometric measurements with bandpasses
indicated by their width.  The red region shows the $1 \sigma$ error on the measurements
while yellow gives the $2 \sigma$ error.  Yellow triangles show
upper limits to narrow band fluxes.
In both panels, spectral intervals are labeled with the
molecules primarily responsible for the opacity in that interval.

\medskip
\item{\bf Fig. 2}  Limits on $T_{\rm eff}$ and $g$ of Gl 229 B.  
The grey shaded area delimits the effective temperature and gravity of model
objects which match within $2 \sigma$ the observed bolometric
luminosity (3) of Gl 229 B at any age.  The other areas show limits 
from fitting the $1-2.5\, \mu$m spectrum (vertical lines) and the $2.5-13\, \mu$m
photometry (horizontal lines).  

\medskip
\item{\bf Fig. 3}
The grey shaded area shows the region of overlap of the three constraints
from Fig. 2 (the  cutoff at low $g$ is arbitrary).
Solid lines depict the evolution of $T_{\rm eff}$ and $g$
as various mass brown dwarfs cool.  Several contours of
constant radius (long-dashed curves) and constant age
(short-dashed curves) are also shown.

\medskip
\item{\bf Fig. 4}
Calculated atmospheric structure for Gl 229 B; the dashed curve shows an
adiabat corresponding to the deep interior temperature profile.  For comparison,
a profile for Jupiter (16) is shown, along with its calculated prolongation
into the adiabatic deep interior (dashed curve).

\vfill\eject
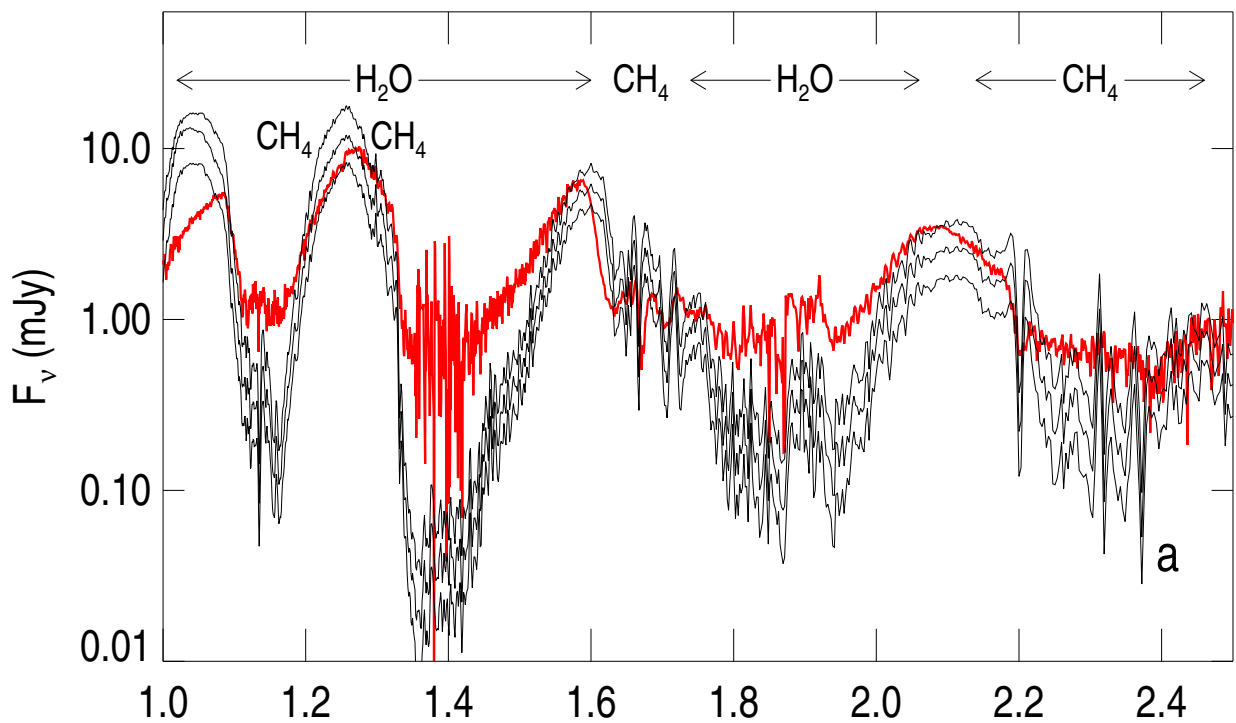
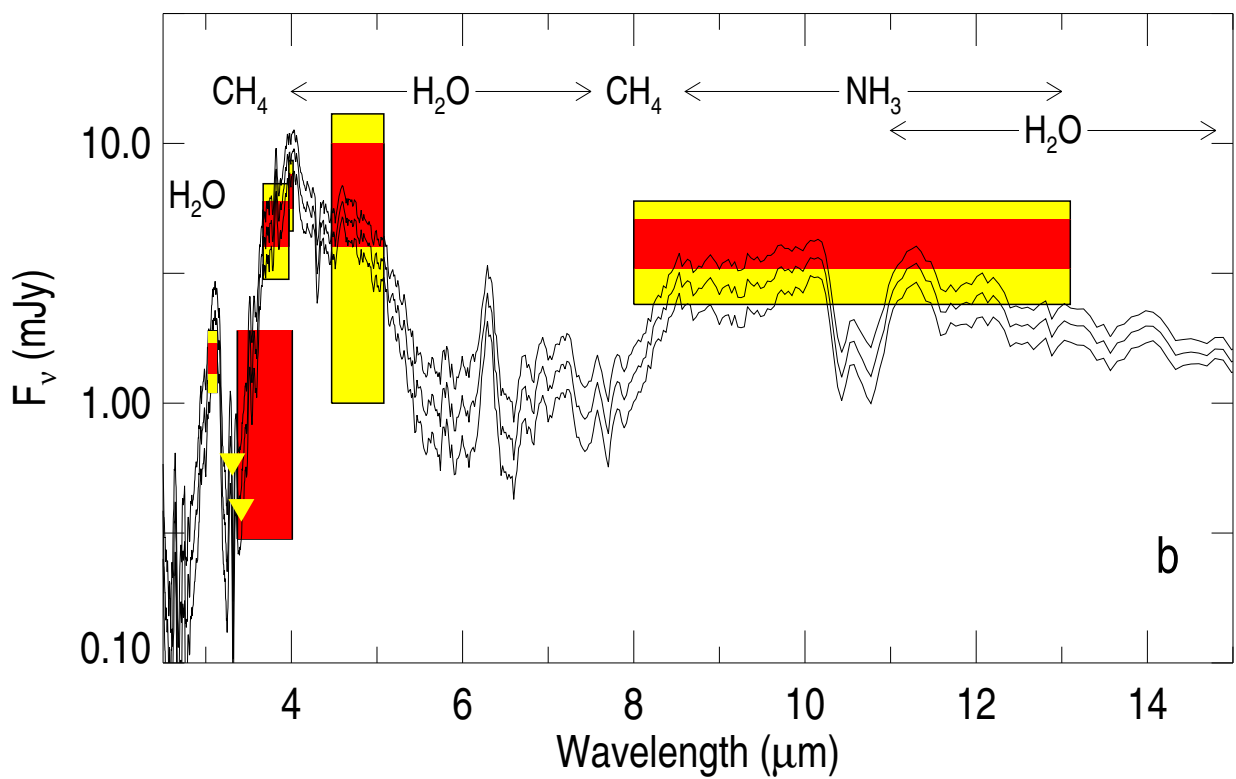
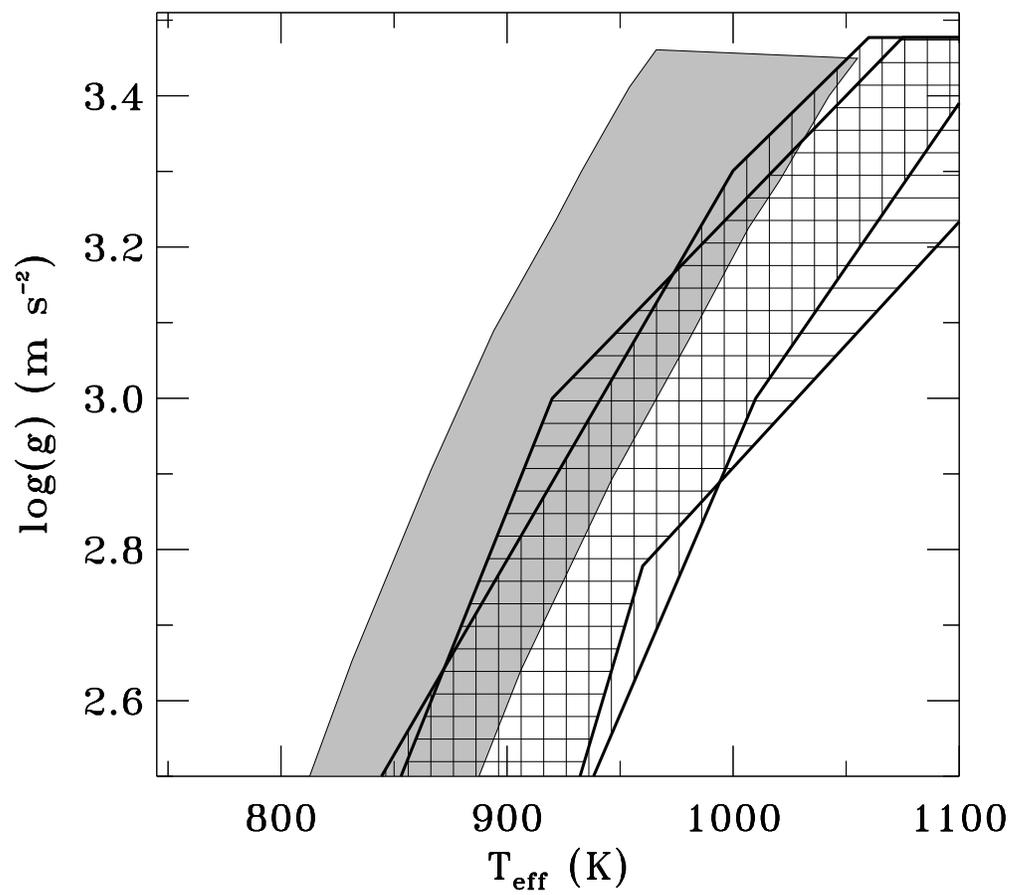
\end